**Constraints for Nuclear Astrophysics from an Unusual Presolar Silicate-Oxide Aggregate Grain Found in Primitive Ordinary Chondrite Meteorite Hills 00526.**

Larry R. Nittler[1] [1], Jens Barosch[2,3], Conel M. O'D. Alexander[2], Jianhua Wang[2]

[1] School of Earth and Space Exploration, Arizona State University, Tempe AZ 85287, USA

[2] Earth and Planets Laboratory, Carnegie Institution of Washington, Washington DC 20015, USA

[3]School of Geosciences, The University of Edinburgh, Edinburgh EH9 3FE, UK

Accepted for publication in *The European Physical Journal A,* May 2, 2026

**Abstract.** We report O, Mg-Al, Si, Ca, and Ti isotopic data for an unusual presolar oxide/silicate aggregate grain, M526-69, previously reported in the primitive ordinary chondrite Meteorite Hills 00526. The ≈1µm aggregate consists of a Mg- and Ca-rich silicate, a Al-rich oxide, and a tiny $TiO_2$ grain. A large $^{18}$O depletion and high inferred $^{26}$Al/$^{27}$Al classifies M526-69 as a Group 2 grain. Both low-mass (LM) and intermediate-mass (IM) asymptotic giant branch (AGB) stars are considered viable candidate parent stars of Group 2 grains based on their O isotopes and inferred $^{26}$Al/$^{27}$Al ratios. The lack of a large $^{30}$Si excess in M526-69 strongly supports an LM-AGB origin for it and other Group 2 grains. The stable Mg, Ca, and Ti isotopes all reflect the initial composition of the parent star, set by galactic chemical evolution (GCE) processes. Presolar O-rich grains provide a better measure of the GCE trends for Ti isotopes than presolar SiC grains as the latter are also affected by neutron capture reactions in the parent stars. Most of the Mg, Ca, and Ti isotopic ratios in M526-69 are consistent with its parent star having metallicity lower than solar. However, small excesses in stable non-radiogenic $^{26}$Mg, $^{46}$Ti, and $^{44}$Ca do not fit this pattern and instead point to heterogeneous GCE processes, though quantitative modeling is needed to test this hypothesis. Multi-phase presolar grains are extremely valuable for nuclear astrophysics as they can both provide isotopic compositions for multiple elements that must be matched at a single time and place in a single star.

---

[1] Email: lnittler@asu.edu

## 1 Introduction

Preserved presolar stardust grains in meteorites are a powerful tool for nuclear astrophysics [1-3]. These tiny (typically <1 μm) and rare (ppb – 100s of ppm) component of primitive meteorites condensed in ancient stellar winds or explosions and became part of the molecular cloud from which the Solar System formed 4.6 Gyr ago. They are recognized as presolar grains because their isotopic compositions reflect those of the stellar gases from which they condensed and these can be strikingly different from those of Solar System materials due to nuclear processes occurring in the parent and prior generations of stars. As "fossilized" bits of stellar matter, they provide important and unique constraints on a wide range of astrophysical and cosmochemical processes.

Many chemical types of presolar grains, both minerals and non-stoichiometric phases, have been identified since the 1987 discovery of presolar SiC, graphite and nanodiamonds [4, 5]. These include a wide range of oxide and silicate phases, including $Al_2O_3$, $MgAl_2O_4$, $CaAl_{12}O_{19}$, $TiO_2$, $(Mg,Fe)_2SiO_4$, $(Mg,Fe)SiO_3$, $SiO_2$, and others. Presolar silicates are the most abundant class of unambiguously presolar grains, reflecting the dominance of silicates in the inventory of interstellar dust [6]. Although most identified presolar grains are dominated by a single phase (though sometimes containing tiny sub-grains of other phases), aggregate presolar grains (also called "complex" grains) consisting of multiple distinct phases stuck together are occasionally found [7-13]. Such aggregate presolar grains are rare: less than 20 have been reported out of >3000 presolar O-rich grains. In general, the isotopic compositions of the individual grains making up the aggregates are very similar, indicating co-origins in the same star; that is, they do not seem to have formed in distinct stars and aggregated later in the interstellar medium. They are highly valuable because distinct chemical phases can allow for a wider range of elements to be analyzed for their isotopic composition and thus provide tighter constraints on the nuclear processes that led to their compositions. Moreover, they can provide detailed constraints on condensation conditions in the parent stellar environments [e.g., 10, 12].

We report here multi-element (O, Mg-Al, Si, Ca and Ti) isotopic data for a compound oxide-silicate presolar aggregate grain, M526-69 (Fig. 1). This aggregate was reported by Barosch et al. [14], who found it consisted of a Mg- and Ca-rich silicate, an Al-rich oxide, and a small Ti-rich grain (most likely $TiO_2$). It has a "Group 2" O-isotopic signature [15], with a 2.6× $^{17}O$ excess and a 60% $^{18}O$ depletion, relative to terrestrial isotopic ratios. Group 2 grains are most commonly believed to have formed in asymptotic giant branch (AGB) stars, with both low-mass and intermediate mass AGB progenitors considered viable [16-18].

Magnesium, Si, Ca, and Ti are all highly useful for presolar grain studies, as their isotopes are affected to varying degrees by internal nuclear processes in parent stars and variations in the initial compositions of stars imprinted by Galactic Chemical Evolution (GCE) processes. Calcium and Ti have proven particularly valuable for studies of presolar C-rich phases like graphite and SiC [19-25], but much less data has been reported for O-rich grains [26-28]. For example, $^{44}Ca$ excesses seen in many presolar SiC and graphite grains are commonly interpreted as the decay product of

$^{44}$Ti and thus as pointing to supernova origins [19, 29]. However, heterogeneity in $^{44}$Ca/$^{40}$Ca ratios in presolar hibonite grains from AGB stars has suggested that GCE may also play a major role [27]. The co-existence of Al-, Ti-, and Mg, Ca, Si-rich phases in M526-69 makes this aggregate valuable for deconvolving GCE and nucleosynthetic effects for a number of isotope systems.

## 2 Methods

Grain M526-69, approximately one micrometer in size, was identified in a polished thin section of the unequilibrated ordinary chondrite (L/LL 3.05) Meteorite Hills (MET) 00526 and reported by Barosch et al. [14]. A field photo from the discovery of this meteorite in Antarctica is shown in Fig. 1a. A secondary electron (SE) image of the grain aggregate is shown in Figure 1b; composite RGB overlays of Ti, Ca, and Al based on scanning-electron-microscopy-based energy dispersive x-ray analysis (EDS) and the NanoSIMS data reported here are shown in Figs.1c and d, respectively.

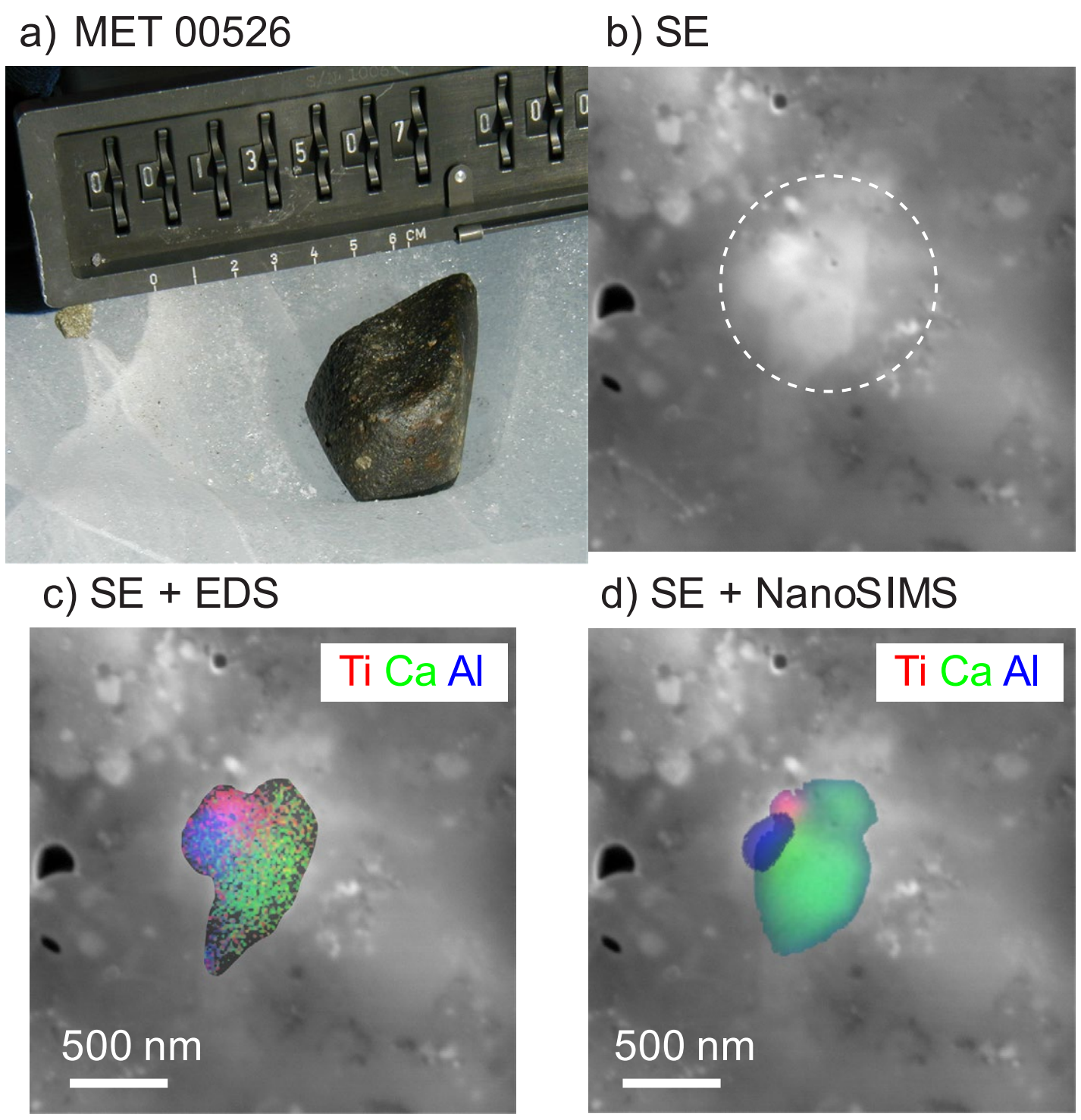

**Fig. 1** a) Field photo in Antarctica of MET 00526 (L. Nittler, January 6, 2001). b) Secondary electron (SE) image of small portion of MET 00526 matrix; the dashed circle indicates the presolar aggregate M526-69. c) SE image overlain with RGB composite showing distribution of Ti, Ca and Al (from SEM-EDS data reported by [14]) in M526-69. d) SE image overlain with RGB composite showing distribution of Ti, Ca and Al (from NanoSIMS data reported here) in M526-69

Previously only NanoSIMS O-isotope and EDS elemental data were reported for M526-69. We further analyzed the aggregate for Mg-Al, Si, K, Ca, and Ti isotopes with the Cameca NanoSIMS 50L at the Carnegie Institution of Washington. An Oregon Physics Hyperion RF plasma source was used to generate a ~100-nm focused primary beam of $^{16}O^-$ ions for analysis of the sample in two analytical sessions. The beam was rastered over the sample with simultaneous collection of positive secondary ions in imaging mode with a pixel resolution of 256×256 and a raster size of 5 μm. In the first session, $^{24,25,26}$Mg, $^{27}$Al, and $^{28,29,30}$Si were analyzed in multi-collection imaging mode. A total of 25 image frames were collected, each integrated for 65.5 s. In the second session, we used combined peak-jumping and multicollection mode to measure K, Ca, and Ti isotopes, using four separate magnetic fields (Fig 2). The total measurement time was about two hours, divided between 28 image frames. Calcium-46, and $^{48}$Ca were of too low abundance to measure and $^{51}$V and $^{52}$Cr were analyzed to estimate potential $^{50}$V and $^{50}$Cr interferences on $^{50}$Ti. Unfortunately, an unidentified interference precluded the accurate measurement of $^{41}$K/$^{39}$K ratios.

The NanoSIMS images were analyzed with the L'image software (L. Nittler, ©Carnegie Institution), including automatic alignment of image frames to correct for stage drift and corrections for electron multiplier deadtime. Isotopic ratios were calculated for manually-defined regions of interest (ROIs) based on elemental and isotopic ratio images. Temporal drift was accounted for in the combined-mode images by linear interpolation of count rates. All isotopic data were normalized to measurements of the surrounding meteorite matrix. The relative sensitivity factors used to convert ion signals to Al/Mg and Ti/Ca ratios were determined via measurements of terrestrial spinel ($MgAl_2O_4$) and perovskite ($CaTiO_3$) grains.

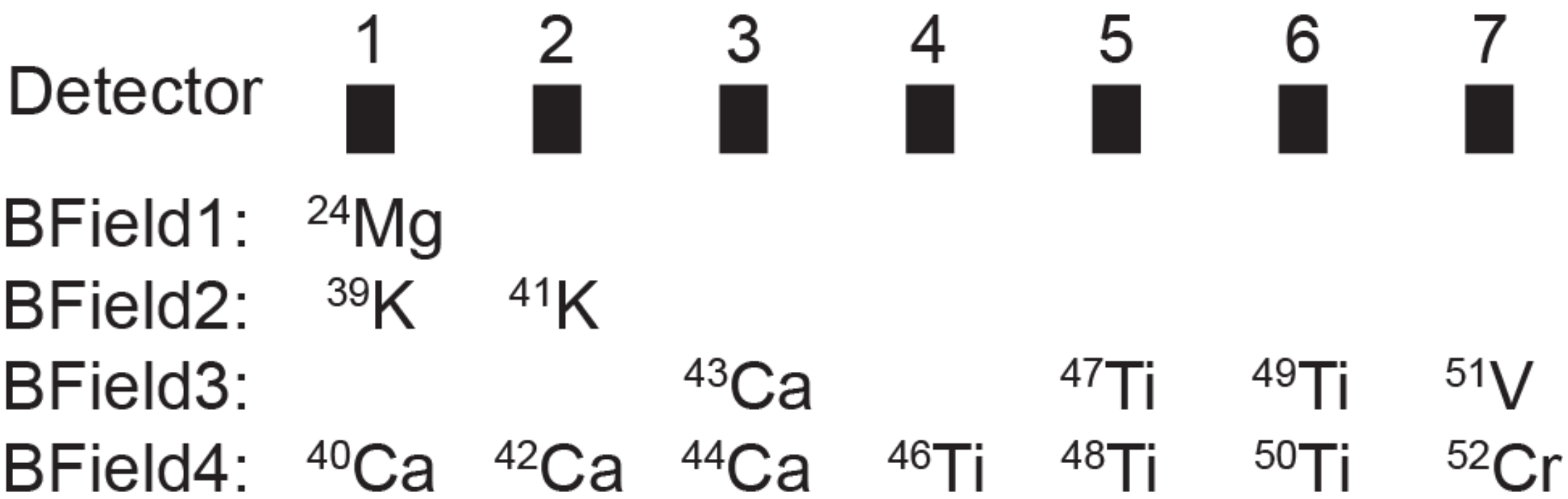

**Fig. 2** NanoSIMS detector setup for K-Ca-Ti measurements

### 3 Results

The new NanoSIMS measurements confirmed that the Al-rich oxide, Ca- and Mg-rich silicate and small $TiO_2$ grain that make up M526-69 are all isotopically anomalous and hence part of the same object. The data for 14 measured isotopic ratios of six elements are summarized in Figure 3 and Table 1.

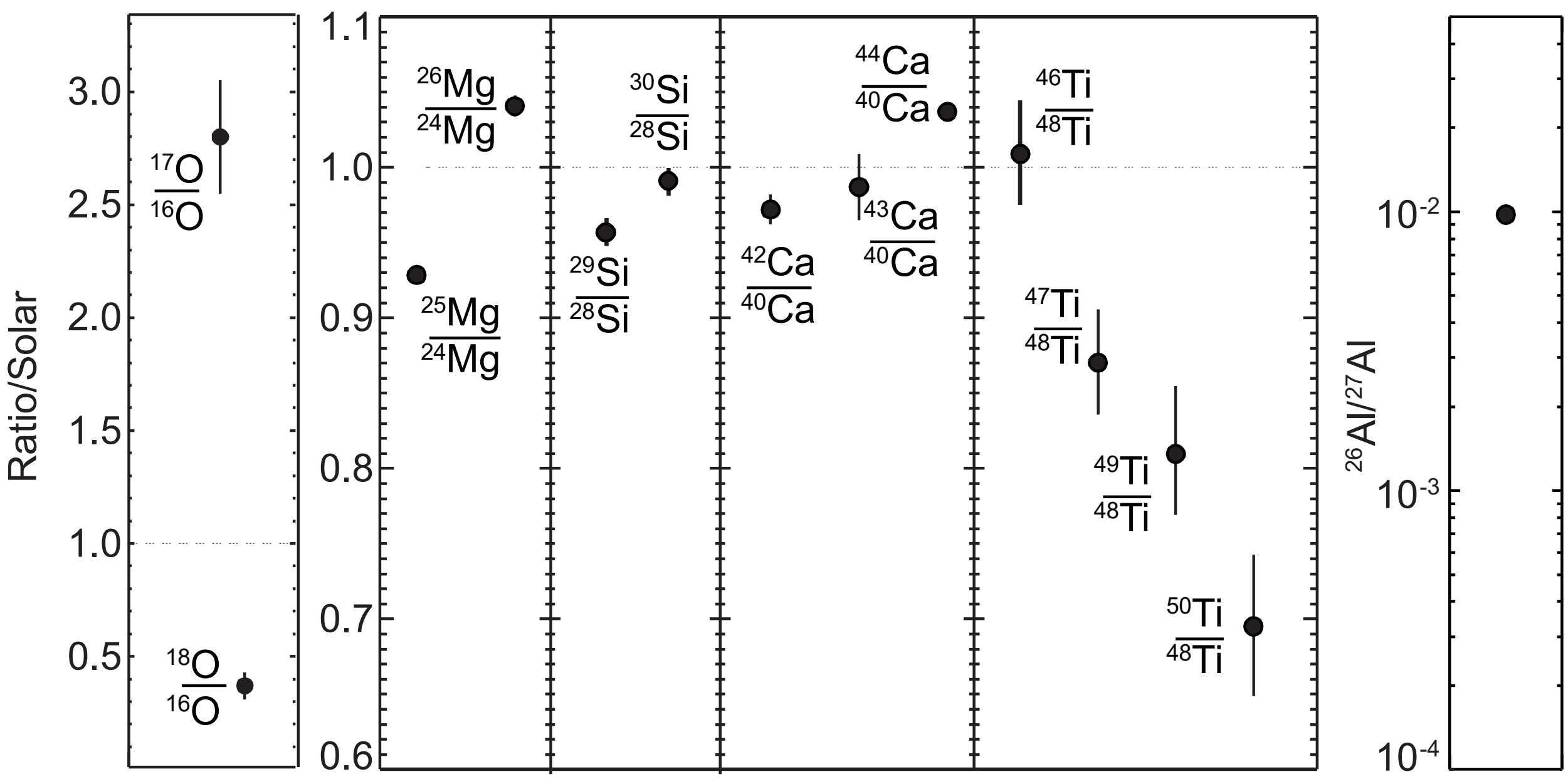

**Fig. 3** Summary of isotopic data for presolar aggregate grain M526-69

**Table 1** Bulk isotopic data for M526-69

| Ratio | Value | 1-σ error |
|---|---|---|
| $^{17}O/^{16}O$ | $1.07 \times 10^{-3}$ | $1.0 \times 10^{-4}$ |
| $^{18}O/^{16}O$ | $7.4 \times 10^{-4}$ | $1.2 \times 10^{-4}$ |
| $\delta^{25}Mg/^{24}Mg$ (‰)[a] | -70 | 6 |
| $\delta^{26}Mg/^{24}Mg$ (‰) | 42 | 6 |
| $^{26}Al/^{27}Al$ (‰) | 0.0098 | 0.0008 |
| $\delta^{29}Si/^{28}Si$ (‰) | -43 | 7 |
| $\delta^{30}Si/^{28}Si$ (‰) | -10 | 8 |
| $\delta^{42}Ca/^{40}Ca$ (‰) | -26 | 10 |
| $\delta^{43}Ca/^{40}Ca$ (‰) | -14 | 20 |
| $\delta^{44}Ca/^{40}Ca$ (‰) | 37 | 6 |
| $\delta^{46}Ti/^{48}Ti$ (‰) | 10 | 35 |
| $\delta^{47}Ti/^{48}Ti$ (‰) | -130 | 35 |
| $\delta^{49}Ti/^{48}Ti$ (‰) | -190 | 40 |
| $\delta^{50}Ti/^{48}Ti$ (‰) | -305 | 40 |

[a]Delta-values: $\delta R$ (‰) $=(R/R_{std} - 1)\cdot 1000$, where R is the measured ratio and $R_{std}$ is the terrestrial ratio.

The O isotopic ratios previously reported for ~3300 presolar silicate and oxide grains are shown in Fig. 4a. The grains are color-coded according to their Group assignments, following

Nittler et al. [15]. As mentioned above, the composition of M526-69 places it in Group 2, with a large $^{17}O$ enrichment and large $^{18}O$ depletion. Although the NanoSIMS images did show some minor O-isotopic variability across the object, this can all be attributed to variable contributions to the measurements of isotopically normal material from the surrounding meteorite matrix.

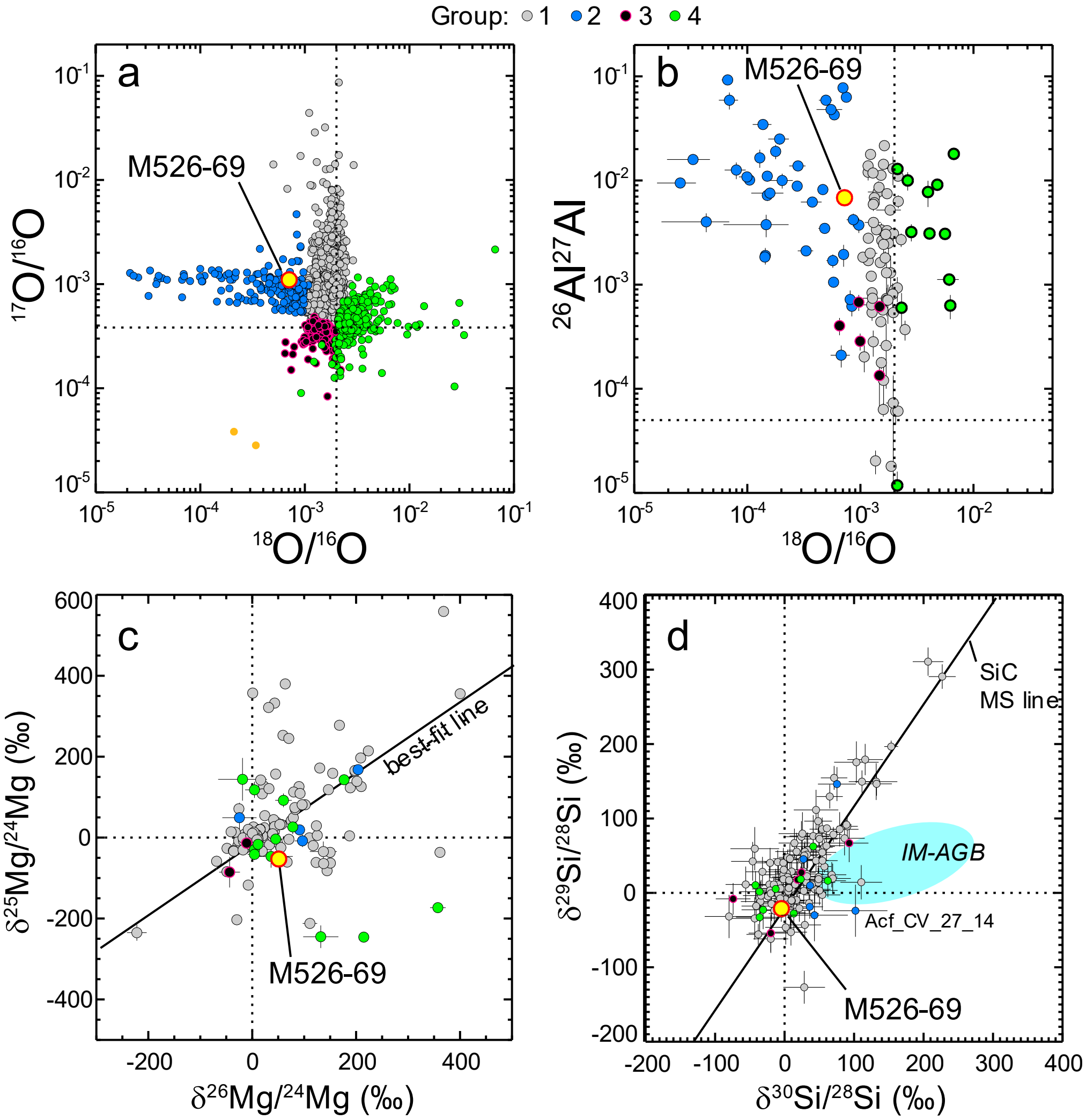

**Fig. 4** Isotopic data for presolar oxide-silicate aggregate M526-69. a) O-isotopic ratios in ~3300 presolar oxide and silicate grains; data are taken from a large number of sources (e.g., [30] and references therein). Grains are color-coded according to their Group assignments [15]. M526-69 is a Group 2 grain. b) Inferred initial $^{26}Al/^{27}Al$ ratios plotted against $^{18}O/^{16}O$ ratios for presolar oxide grains and M526-69. c) Mg 3-isotope plot of presolar silicates and M526-69 [31-33]. Ratios are expressed as δ-values (see Table 1). The black line indicates the best-fit line of Hoppe et al. [32] to "normal" Group 1 grains (see text). d) Silicon 3-isotope plot of presolar silicates [31-33] and M526-69. The solid line indicates the "mainstream (MS)" line defined by the majority population of presolar SiC grains [34]. The cyan ellipse indicates the predicted envelope composition for a 5M⊙ AGB star [35]. Group 2 grain Acf_CV_27_14 was reported by Hoppe et al. [32]. Dotted lines in all panels indicate terrestrial isotopic ratios (or early solar System value for $^{26}Al/^{27}Al$)

The Mg-Al isotopic systematics of M526-69 are shown in Fig. 5. The distinct Al/Mg ratios of the grains making up the aggregate allowed a mixing line to be constructed (Figs. 5b,c). The slope of this isochron-like line indicates that the grain condensed with a $^{26}Al/^{27}Al$ ratio of ≈0.01 (Fig. 4b), a typical Group 2 signature. The intercept indicates that the non-radiogenic, initial $\delta^{26}Mg$ value of the grain was 42±6‰ (Fig. 4c). The $^{25}Mg/^{24}Mg$ ratio was found to be depleted, relative to terrestrial, with $\delta^{25}Mg$= -70±6‰ (Figs. 4c, 5a).

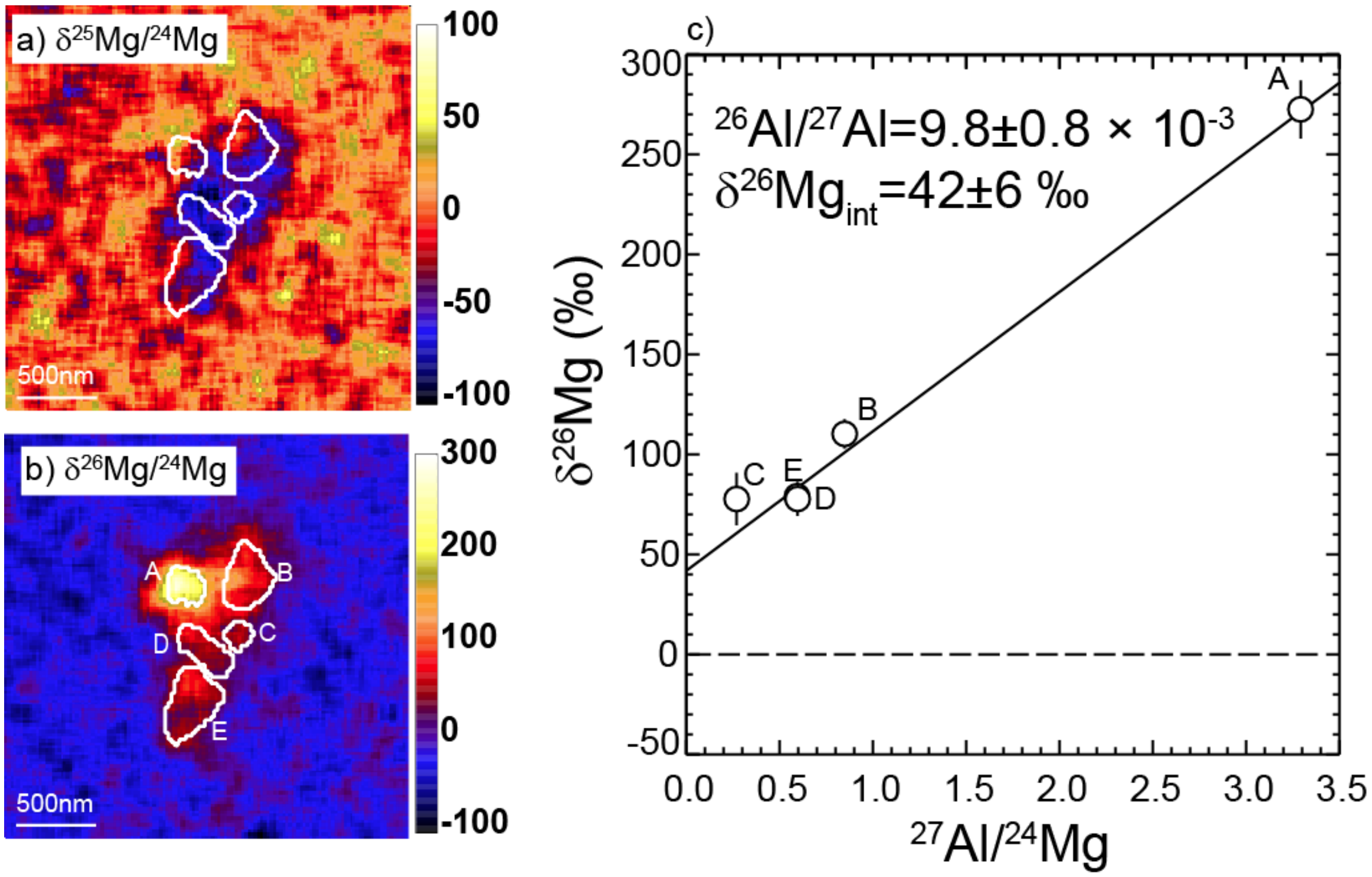

**Fig. 5** Al-Mg isotope systematics of M526-69. a) NanoSIMS $^{25}Mg/^{24}Mg$ and b) $^{26}Mg/^{24}Mg$ image. White outlines indicate five manually defined regions of interest (ROIs). c) Al-Mg "isochron" plot shows that the ROIs are linearly correlated, allowing determination of both the initial $^{26}Al/^{27}Al$ ratio and initial non-radiogenic $^{26}Mg/^{24}Mg$ ratio. The former is determined from the slope of the best-fit line by first converting $\delta^{26}Mg$ back to $^{26}Mg/^{24}Mg$: $^{26}Al/^{27}Al = [(^{26}Mg/^{24}Mg)_{meas} - (^{26}Mg/^{24}Mg)_{\odot})]/ [^{27}Al/^{24}Mg]$

The heavy Si isotopes were found to be slightly depleted in the silicate portion of the grain, with $\delta^{29}Si$= -43±7‰; $\delta^{30}Si$ = -10±8‰ (Fig 4d). There was insufficient Si in the Al-rich and Ti-rich grains to determine their Si-isotopic composition. This Si-isotopic composition places M526-69 at the low end of the line defined by the dominant "mainstream" population of presolar SiC grains [34].

The Ca isotopic pattern of M526-69 (Fig. 6a) is very similar to those of two previously reported presolar hibonite ($CaAl_{12}O_{19}$) grains KH2 (Group 4) and KH13 (Group 2) grains [27], namely it has a small depletion in $^{42}Ca$ and small excess in $^{44}Ca$ (Table). Calcium-44 excesses in presolar grains are often attributed to *in situ* decay of $^{44}$Ti and thus seen as an indicator of a supernova origin. The presence of both Ti-rich and Ca-rich grains in this aggregate allows us to test this interpretation for M526-69. In fact, both grains show identical $\delta^{44}$Ca values within error. Based on the measured Ti/Ca ratios, if the (~4%) $^{44}$Ca excess in the silicate grain was due to $^{44}$Ti decay, the inferred initial $^{44}$Ti/$^{48}$Ti ratio would be 0.017, With this much $^{44}$Ti, the Ti-rich grain would be expected to show a $^{44}$Ca/$^{40}$Ca ratio some 2.5 times higher than solar (star symbol in Fig 6b). We therefore conclude that the $^{44}$Ca enrichment in this aggregate cannot be explained by $^{44}$Ti decay.

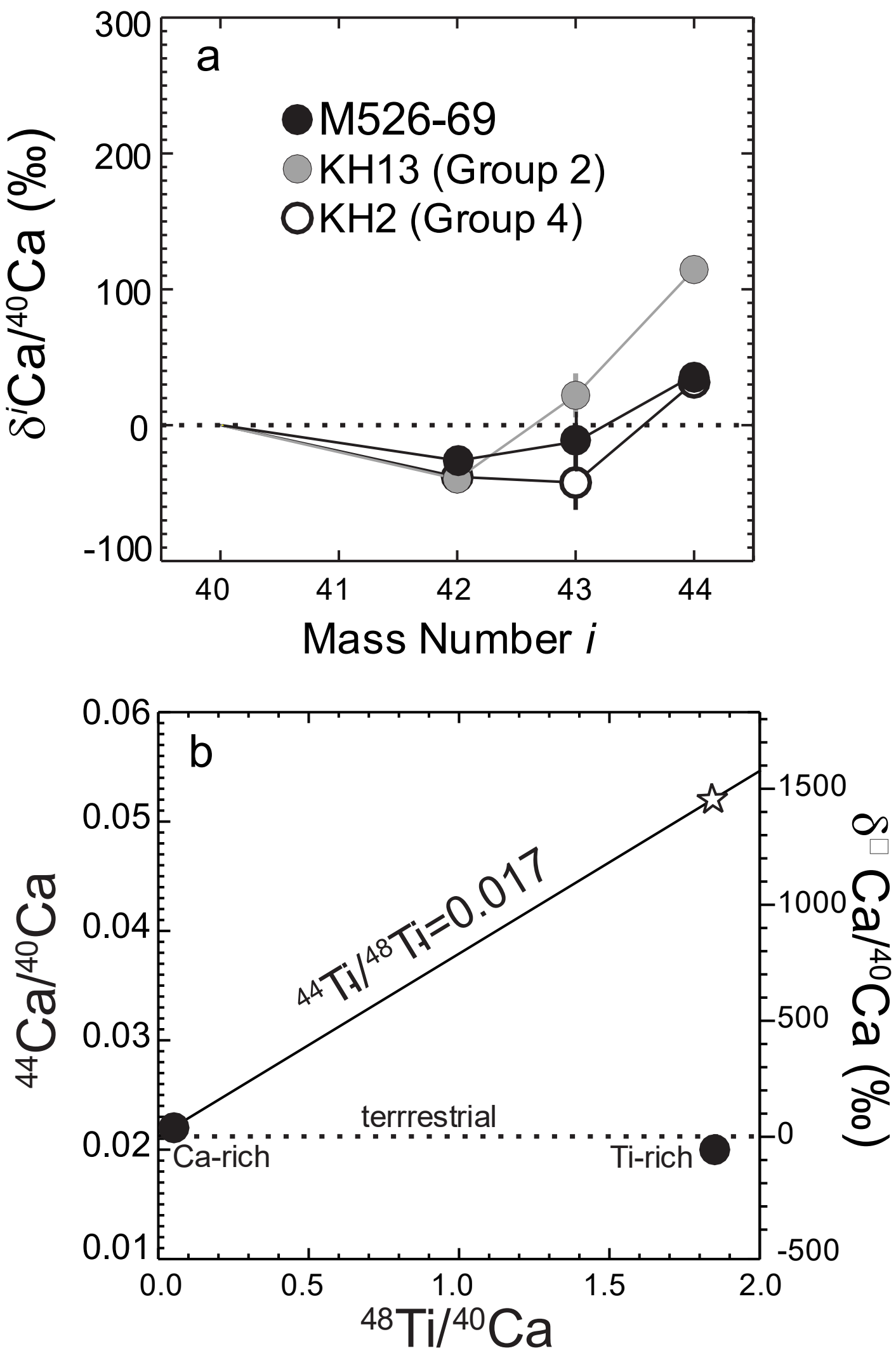

**Fig. 6** a) Ca isotopic ratios (expressed as δ-values) of M526-69 are compared to two previously reported presolar hibonite grains [27]. b) The $^{44}$Ca/$^{40}$Ca ratios is plotted against the $^{48}$Ti/$^{40}$Ca ratio for the Ca-rich and Ti-rich sub-grains of aggregate M526-69. If the small $^{44}$Ca excess observed in the Ca silicate was due to in-situ decay of $^{44}$Ti, the Ti-rich sub-grain would be expected to have a much higher $^{44}$Ca excess (star symbol). We thus conclude that the $^{44}$Ca excess is not radiogenic

The Ti isotopic measurement revealed a normal $^{46}Ti/^{48}Ti$ ratio, but 13-30 % depletions in $^{47}Ti$, $^{49}Ti$, and $^{50}Ti$ (Table1, Fig. 3). The composition was identical within errors for both the Ti-rich sub-grain and the larger Ca-rich silicate; there was insufficient Ti in the Al-rich grain to measure.

## 4 Discussion

### 4.1 Origins of Group 2 O-rich Presolar Grains

Presolar O-rich grains have been historically divided into four main groups (with a few additional outliers) based on their O isotopic ratios [15]. Group 1, the dominant population, is characterized by moderate to large enrichments in $^{17}O$ and $^{18}O/^{16}O$ ratios ranging from about half terrestrial to about terrestrial. These compositions are in good agreement with predictions for low-mass (roughly less than 3-4 solar mass) AGB stars and reflect the mixing of ashes of core H-burning into the convective envelope during the first dredge-up process upon ascent of the first red giant branch. Such stars are, thus, commonly accepted as the main source of Group 1 grains, though a still uncertain fraction of them may form in red supergiants and/or Type II supernovae [27, 36]. The first dredge-up has only a minor effect on $^{18}O/^{16}O$ ratios, so the range of this ratio observed in Group 1 grains is thought to reflect variations in the starting compositions of the parent stars, induced by GCE processes. Group 3 grains have been argued to originate in low-mass, low-metallicity[2] AGB stars [15] and the $^{18}O$-enriched Group 4 grains in Type II supernovae [26, 27].

As mentioned earlier, Grain M526-69 is a Group 2 grain in terms of its O isotopes and inferred $^{26}Al/^{27}Al$ ratio (Figs. 4a,b). It was recognized very early that while the $^{17}O/^{16}O$ ratios of Group 2 grains are also consistent with AGB stars, their $^{18}O$ depletions (Figure 4a) are too large to be explained by the first dredge-up. Instead, they require $^{18}O$+p reactions to affect the stellar envelope material itself. Two processes have been suggested to explain this nuclear burning. First, in intermediate-mass AGB (IM-AGB) stars (4–7 $M_{\odot}$), "hot-bottom burning" is expected to occur, where the base of the stellar envelope reaches temperatures sufficient for proton-capture reactions to take place [37]. In this environment, convective envelope material is cycled through the hot region where $^{18}O$ gets rapidly destroyed. Conversely, in low-mass AGB (LM-AGB) stars ($<\sim1.5$ $M_{\odot}$), an extra mixing process may occur, termed "cool bottom processing," [17, 38, 39] in which envelope material is non-convectively transported to hotter, deeper regions where it undergoes nuclear burning, before being cycled back to the envelope itself. The H-burning reactions taking place in either process will produce $^{26}Al$ as well. The physical cause of the mixing in cool-bottom processing is still unknown, though the most successful model so far involves buoyant ascent of magnetized bubbles within the star [40, 41].

For many years, an origin for Group 2 grains in IM-AGB stars was not considered viable because hot-bottom burning models predicted substantially higher $^{17}O/^{16}O$ ratios than observed in the grains [37]. This changed with a new direct measurement of the rate for the critical reaction

[2] *Metallicity* (or *Z*) is the mass fraction of elements heavier than helium in an astronomical environment; solar metallicity is between 1.5 % and 2 %.

$^{17}O(p,\alpha)^{14}N$, which indicated more efficient destruction of $^{17}O$ and hence lower $^{17}O/^{16}O$ ratios than previously calculated for IM-AGB stars [16]. Recently Palmerini et al. [17] reported new calculations for both LM-AGB stars with magnetically-induced cool-bottom processing and IM-AGB stars with hot-bottom burning, considering two different sets of nuclear reaction rates. These authors favored a low-mass AGB origin for the grains but concluded that remaining uncertainties in both the nuclear physics and the stellar models do not allow IM-AGB parent stars to be ruled out. One difficulty with an IM-AGB origin is that the hot-bottom burning leads to essentially complete destruction of $^{18}O$, so that an additional mixing episode must be invoked to explain the broad range of $^{18}O/^{16}O$ ratios observed in the grains, whereas the cool-bottom processing models naturally explain this range.

One difficulty in resolving the low-mass versus intermediate mass AGB star origin of Group 2 grains has been that only O isotopes and inferred $^{26}Al/^{27}Al$ ratios have been available for most grains (mainly Al-rich oxide phases) and these can be explained in both scenarios. Group 2 presolar silicates may help distinguish these models. Note that IM-AGB star envelopes are predicted to have large excesses in $^{30}Si$, due to mixing of material experiencing neutron capture reactions in the intershell region between the H- and He-burning shells during thermal pulses and third dredge-up episodes [35, 42]. In fact, very few Group 2 silicates have been reported, in large part reflecting the fact that presolar silicates are found in situ in meteorite slices and their isotopic measurements can be severely affected by contamination by surrounding material. Nevertheless, Si data are now available for a few such grains (Fig. 4d), including the M526-69 data reported here, and the grains are not generally $^{30}Si$-rich compared to Group 1 grains. As discussed further in the next section, M526-69 lies on the Si-isotope line defined by the majority population ("mainstream") of presolar SiC grains and most presolar silicates. This indicates that its parent star did not mix material that experienced neutron capture reactions into its envelope and strongly supports an origin for this Group 2 aggregate in a LM-AGB star that underwent cool bottom processing. We note that one Group 2 silicate grain (Acf_CV_27_14) [32] does lie to the $^{30}Si$-rich side of the mainstream line (Fig. 4d) and thus might be indicative of a more massive parent star, but the statistical significance is not high.

**4.2 Mg and Si isotopes in M526-69**

Magnesium shows a wide range of isotopic compositions in O-rich presolar grains. Aluminum-rich oxides very commonly exhibit larges excesses in $^{26}Mg$ from in situ decay of $^{26}Al$ ($t_{½}$ =720,000 yr). Magnesium-rich phases like spinel ($MgAl_2O_4$) and many silicates allow nucleosynthetic effects on Mg itself to be investigated [27, 29, 31, 36, 43]. Unlike most major elements in presolar grains, a significant fraction of both presolar spinel and silicate grains have close-to-solar Mg isotopic ratios. This has been suggested to have arisen from isotopic exchange in space or within meteorite parent bodies [27, 31], but this remains to be proven. Many grains do show anomalous Mg isotopes, however. Hoppe et al. [31, 32] recently reported Mg and Si isotopic ratios for >100 presolar silicate grains (Figs. 4c,d) and showed that the dominant Group 1 population can be further sub-divided into sub-groups based on distinct Mg isotopic properties. About 60% of Group 1 grains with resolvable Mg isotopic anomalies plot along a line on the Mg 3-isotope plot (Fig. 4c) with a slope of ~0.9, termed "normal" grains. The other anomalous Group 1 grains fall either above ("$^{25}$Mg-rich") or below (divided into two groups: "$^{26}$Mg-rich" and "$^{25}$Mg-poor") this line and were interpreted by Leitner and Hoppe [36] and Hoppe et al. [32] to indicate an origin in massive stars and/or Type II supernovae. The few Group 2, 3, and 4 silicate grains with Mg anomalies showed a similar range of Mg isotopes to the Group 1 grains. The Mg-isotopic composition of Group 2 grain aggregate M526-69 (Fig. 4c) places it near the lower end of the silicate line defined by the normal Group 1. It is displaced to a slightly $^{26}$Mg-rich composition, relative to this line, but it is still in the range of compositions defined by Hoppe et al. as "normal Group 1."

Thus far, Si isotopic ratios have shown a smaller range of anomalous compositions among presolar silicates compared to Mg isotopes. Most grains scatter around a line on the Si 3-isotope plot of slope ~1.3 (Fig. 4d). This linear array is also what has been long observed for Si isotopes in "mainstream" presolar SiC grains, which make up some 90 % of presolar SiC and are well-established to have originated in AGB stars of near-solar or super-solar metallicity [35]. The slope of this line has long been recognized to be much steeper than can be explained by neutron capture reactions in the parent stars, and thus is interpreted to reflect the initial compositions of the parents, set by GCE processes [44]. The Si isotopic ratios in presolar silicate grains can clearly be interpreted in the same manner, indicating that they formed from a similar population of stars as birthed the SiC grains. Moreover, the "normal" Group 1 grains defined by Hoppe et al. [32] show a strong linear correlation between $^{25}$Mg/$^{24}$Mg and $^{29}$Si/$^{28}$Si ratios, indicating that the Mg isotopic variations in these grains reflect variations in the parent stars' initial compositions as well.

GCE is the theory that describes how the chemical composition of galaxies changes over time as new generations of stars are born, evolve, and inject newly-synthesized elements into the interstellar medium. Isotopic ratios are expected, and observed in some cases, e.g. [45-47], to vary in time and place since the different isotopes are made in different types of stars, by different processes, and over different timescales. We distinguish between *primary* isotopes, whose nucleosynthetic yields are independent of metallicity, and *secondary* ones, whose synthesis does depend on the initial metallicity of the star. GCE modeling shows that the ratio of a secondary isotopes to a primary one should be roughly linearly proportional to metallicity and increase over time in the Galaxy [48, 49]. For Mg and Si, the most abundant isotopes (e.g., $^{24}$Mg, $^{28}$Si) are primary, while the minor isotopes are secondary. Therefore, at least to a first approximation and assuming the isotopes are not modified by nuclear processes in the parent stars themselves, a grain showing lower isotopic ratios of these elements most likely originated in a star of lower metallicity than the parent star of a grain with higher isotopic ratios. We note however, that GCE is not a simple, uniform process, and chemical and isotopic heterogeneities are expected due to the stochastic nature of star formation. That is, while these isotopic ratios are expected to have increased on average in the Galaxy, we also would expect scatter around average GCE trends [50, 51]. In any case, the location of grain M526-69 near the bottom of the Mg and Si isotopic arrays suggests that it formed from a relatively low-metallicity (sub-solar) low-mass AGB star.

### 4.3. Titanium Isotopes in M526-69

Like Mg and Si, Ti isotopic ratios are expected to vary with metallicity due to GCE [52]. However, the nucleosynthetic yields of all five stable Ti isotopes are highly uncertain, leading to very large uncertainty in GCE predictions. For example, predictions from two different GCE calculations [52, 53], shown in Fig. 7, give strongly different results, with neither reproducing the solar ratios at solar metallicity (Fig. 7). Both GCE calculations underproduce the solar Ti elemental abundance by about a factor of two as well. Chavez and Lambert [47] reported Ti isotopes measured spectroscopically in eleven Milky Way M dwarf stars and found that all had ratios that are similar to solar, despite spanning one order of magnitude range in metallicity (Fig. 7). However, the measurement uncertainties are very large and do not allow for distinguishing between GCE models. In this regard, the much higher precision data obtained for presolar grains may be of great importance toward better understanding the galactic history of Ti isotopes.

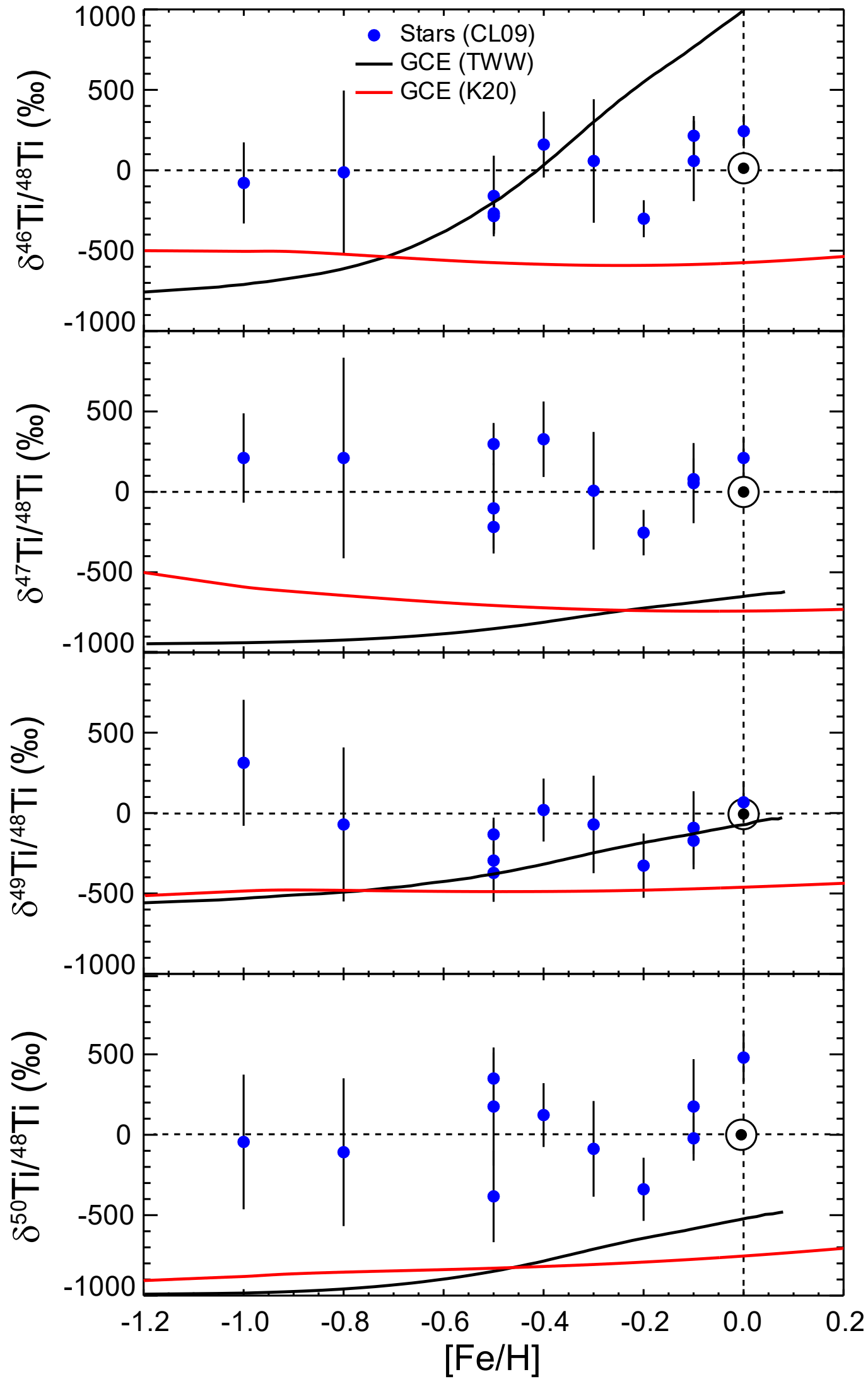

**Fig. 7** Observational and theoretical studies of the Ti isotopic evolution of the galaxy. Blue symbols are spectroscopic measurements of 11 M dwarf stars [47]. The black curve shows the predicted evolution of the Galaxy from Timmes et al. [52], while the red curves show that from Kobayashi et al. [53]

The Ti isotopic composition of M526-69 is compared with those of AGB-derived presolar SiC, $Al_2O_3$, and $TiO_2$ grains in Fig. 8. Titanium isotopic systematics have been well studied in AGB-derived presolar SiC [23, 25, 54, 55]. The $^{46}Ti/^{48}Ti$, $^{47}Ti/^{48}Ti$, and to a lesser extent $^{49}Ti/^{48}Ti$ ratios are well correlated with each other and with $^{29}Si/^{28}Si$ ratios in both mainstream SiC and the rare (few %) Z SiC sub-group (Fig 7). Z grains are commonly interpreted as having formed in low-metallicity, low-mass AGB stars [35]. Although Ti isotopic data for presolar oxides are very limited, the available data for these grains seem to follow the same basic trends as the SiC grains. The Si-Ti correlations strongly support the interpretation that, as for Si and Mg, these isotopes reflect, at least in part, the initial compositions of the parent stars, set by GCE processes. However, neutron capture reactions in AGB stars can also produce Ti isotopes, especially $^{49}Ti$ and $^{50}Ti$, which can be mixed into the envelope during third dredge-up episodes. Thus, the correlation trends are poorer for $^{49}Ti/^{48}Ti$ and $^{50}Ti/^{48}Ti$ (Figs. 8c,d), since the SiC parent stars' initial compositions have been overprinted with newly synthesized Ti isotopes. The presolar oxides likely formed prior to many third dredge-up episodes (which also mix $^{12}C$ to the surface, eventually precluding O-rich dust condensation) and therefore are better indicators for the parent stars' initial Ti isotopic compositions. We thus estimate the GCE trends for the Ti isotopes based on best-fit lines to the SiC (Fig. 8b) or oxide (Figs. 8c,d) data (solid lines).

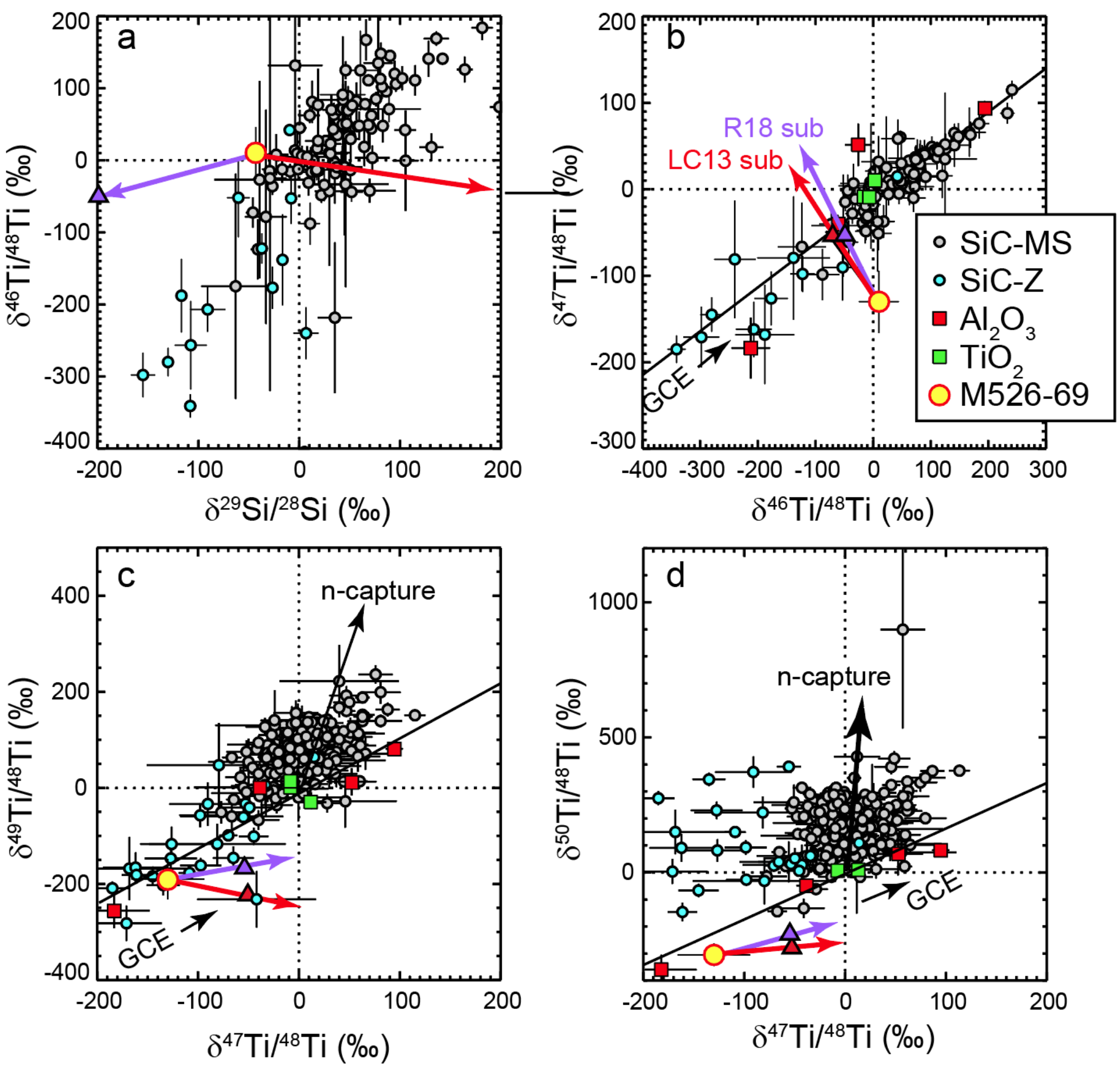

**Fig. 8** Silicon- and Ti-isotopic systematics of M526-69 are compared with those of presolar SiC, $Al_2O_3$, and $TiO_2$ grains [23, 25-27, 54, 55]. Only the majority "mainstream (MS)" and rare Z-type SiC grains are plotted as these are accepted to have formed in low-mass AGB stars. Solid lines are weighted linear fits to the SiC (b and c) or oxide (d) data and approximate the inferred GCE trends for these isotope ratios. M526-69 plots close to these trends with the exception of its $^{46}Ti/^{48}Ti$ ratio. Dotted lines indicate the solar isotopic ratios. The black solid arrows indicate directions of GCE and mixing of neutron-capture material in AGB stars. Red and purple arrows and triangles indicate subtraction of average supernova material from grain M526-69 composition based on yields from Limongi and Chieffi ["LC13", 58] and Ritter et al. ["R18", 59], respectively; see text for details.

For most of the ratios plotted in Fig. 8, grain aggregate M526-69 lies near the GCE Ti-isotopic trends defined by the prior SiC and oxide data, near the SiC Z grains and one previously reported presolar $Al_2O_3$ grain. This provides additional strong support for the interpretation that this aggerate formed in a star of lower-than-solar metallicity (e.g., Z grains have been suggested to come from AGB stars as low as one third of solar metallicity, [56]). However, the ~solar $^{46}Ti/^{48}Ti$ ratio in M526-69 is notably discrepant. For example, based on its $\delta^{47}$Ti value, to fall within the region of Z-grains on Fig. 8b, the grain would be expected to be depleted in $^{46}$Ti by 100 ‰ to 200 ‰. Given that the other Ti isotopes as well as Mg, Si, and Ca (see below) isotopes all support a low-metallicity origin for the grain, it is more likely that the parent star had an unusual $^{46}$Ti/$^{48}$Ti ratio than that the GCE interpretation is incorrect. One plausible explanation is that it reflects heterogeneous GCE. Lugaro et al. [50] first suggested that the Si-isotopic spread in mainstream SiC might reflect stochastic variations in the input of supernova material into different star-forming regions where the parent stars formed. Nittler [51] expanded this model to consider Ti isotopes and showed that the good correlation between $^{29}$Si/$^{28}$Si and $^{46}$Ti/$^{48}$Ti ratios could be used to constrain the degree of heterogeneous GCE that influenced the parent stars of mainstream SiC grains. He found that, within the assumptions of the model and the specific supernova nucleosynthetic yields used, up to about 100 ‰ range in $^{46}$Ti/$^{48}$Ti ratios might be expected from stochastic mixing of supernova ejecta into an initially homogeneous composition. Examination of Fig. 8b shows that the isotopically light SiC grains show a similar degree of scatter around the trend line. Grain 526-69 lies slightly further away from the average trend, but not remarkably so given the magnitude of the uncertainties on the isotopic measurements.

The model of Nittler [51] was based on modified Type II supernova yields from Rauscher et al. [57]. To further test the idea that the slightly discrepant $^{46}$Ti/$^{48}$Ti ratio of grain M526-69 is due to heterogeneous GCE, we considered two more recent sets of yields from Limongi and Chieffi [58] and Ritter et al. [59]. We assumed that the parent star formed from a molecular cloud that had experienced the admixture of a small amount of additional supernova material compared to the average cloud of its age. A full Monte Carlo model of the sort reported by Nittler is out of scope of this paper. Instead, we calculated the average yields of supernovae from the two model sets, weighted by a Salpeter initial mass function, for Si, Ca, and Ti isotopes. We then investigated the effect of subtracting small fractions of these average yields from the initial composition of the parent star of M526-69, assumed to be solar but with the rare isotopes adjusted to match the grain measurements. The resulting trajectories on the Si-Ti three-isotope plots are indicated with colored arrows on Fig. 8, with the colored triangles indicating the effect of subtracting 0.2% (Limongi and Chieffi) or 0.3% (Ritter et al.) of the IMF-weighted supernova yields (for solar metallicity and masses $\leq 25 M_{\odot}$). Fig 8b shows that with this amount of subtracted supernova material, the initial composition of the grain's parent star moves onto the average GCE trend derived from presolar SiC grain data. The same subtraction moves the other Ti isotopic ratios (Fig. 8c,d) away from the presolar grain-derived GCE trends, but not far outside the range of the SiC and oxide data. This strengthens the case that heterogeneous mixing of supernova material in the interstellar medium is a likely explanation for the grain data. However, it must be noted that there are strong uncertainties in the yields, reflected in this case by the strongly diverging predictions for the two models for Si isotopes (Fig. 8a), with the subtraction of the Ritter et al. [59] yields predicting an initial composition strongly depleted in $^{29}$Si and the Limongi and Chieffi [58] yields giving strongly positive $^{29}$Si (off-scale at $\delta^{29}$Si=+260 ‰). The main difference between the two sets of supernova calculations is that some of the Ritter et al. models experienced mergers of convective O- and C-shells, which strongly impacted the nucleosynthesis [60]. Examining the yields in detail, we find that for the supernova masses considered, the Si, Ca, and Ti isotopes vary on average between the two model sets by a factor of about 2.5, but with some yields differing between the models by as much as a factor of 26. Therefore, the results of our simple mixing model are highly uncertain and we defer more detailed quantitative consideration of how heterogeneous GCE affects isotopic ratios to future work.

### 4.4 Calcium isotopes in M526-69

The Ca isotopic composition of aggregate M526-69 is compared in Fig. 9 to previously reported data for presolar hibonite ($CaAl_{12}O_{19}$) grains [27, 61]. Also shown are the ranges predicted for the plotted isotopic ratios in low-mass (1.5 - 3 $M_\odot$) AGB stars of solar and half-solar metallicity, calculated by Roberto Gallino and reported in Nittler et al. [27]. These represent the range of compositions for the stars while their envelopes are O-rich (O>C), required to condense oxide and silicate phases. The predicted evolution of the Ca isotope ratios from the GCE model of Timmes et al. [52], normalized to pass through the solar composition, are also shown. The half-solar metallicity model stars were assumed to have model-dependent $^{40}$Ca-rich starting compositions that, in the case of the $\delta^{42}$Ca versus $\delta^{44}$Ca plot (Fig. 9b), are far from the predictions of the GCE model. However, the plots clearly show that s-process nucleosynthesis in low-mass AGB stars is not predicted to cause large changes in the envelope Ca isotopic composition while the stellar envelopes are O-rich, even at low metallicity. Since the spread in the presolar grain data is clearly larger than expected for AGB evolution, as seen before for Mg, Si, and Ti, the ratios most likely reflect the initial compositions of the grains' parent stars. On the $\delta^{42}$Ca versus $\delta^{43}$Ca plot (Fig. 9a), most of the hibonite grains are well correlated and lie close to the predicted GCE trend. M526-69 lies near the lower end of the trend, again supporting a relatively low metallicity for its parent star.

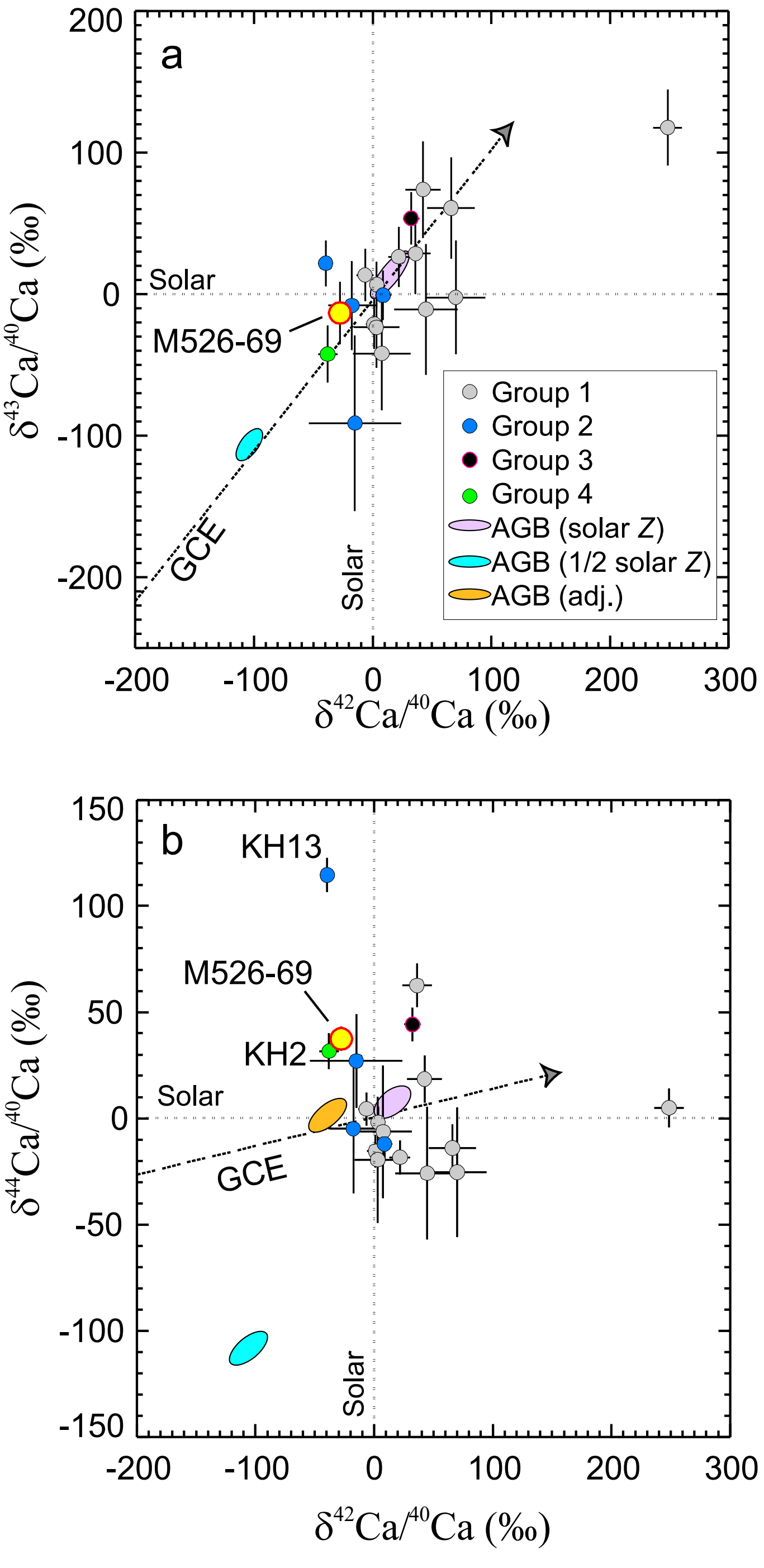

**Fig. 9** Calcium isotopic data for presolar hibonite grains [27, 61] and oxide-silicate aggregate M526-69. Colored ellipses indicate predictions for the envelopes of 1-3 $M_{\odot}$ AGB stars of solar and half-solar metallicity [27]. The dashed arrows indicate the solar-normalized GCE trends calculated by Timmes et al. [52].

The $\delta^{42}$Ca versus $\delta^{44}$Ca plot (Fig. 9b) tells a different story, however; the grain data are uncorrelated. As mentioned above, the half-solar metallicity AGB models were assumed to have initial compositions far below the predicted GCE trend, leading to a large offset between those model predictions and the grain data. The "AGB (adj.)" ellipse on Fig. 9b shows our estimate of the range of O-rich AGB outflows for a lower-than-solar metallicity star with initial Ca-isotope composition on the predicted GCE trend. Several grains from all four O isotope Groups, including Group 2 M526-69, are enriched in $^{44}Ca/^{40}Ca$, compared to the expected GCE trend and small variations expected from *s*-process nucleosynthesis. Nittler et al. interpreted this excess in Group 4 grain KH2 as being radiogenic from decay of (60-yr $t_{½}$) $^{44}Ti$. However, as discussed above in Section 3, for M526-69 this can be unambiguously ruled out since the Ti-rich sub-grain does not exhibit an extreme $^{44}Ca$ enrichment (Fig. 6) and is unlikely for the other non-Group 4 grains as well. We note that new $^{42}Ca/^{40}Ca$ and $^{44}Ca/^{40}Ca$ data for several presolar silicates have been reported graphically in recent conference abstracts [e.g., 28] and these show a very similar distribution. The main nucleosynthetic source of $^{44}Ca$ in the Universe is $^{44}Ti$, which is generally predicted to form in abundance in Type II supernovae. However, based on a lack of detected gamma-rays from $^{44}Ti$ decay from the inner Galaxy, The et al. [62] argued that perhaps only rare supernovae synthesize $^{44}Ti$. These authors further suggested that a testable prediction of this would be large (factor of 2) scatter in $^{44}Ca$ abundances in presolar grains. This is much larger than observed in the presolar hibonite data, so the grain data do not seem to support this suggestion. Moreover, recent multi-dimensional Type II supernova models [63] predict $^{44}Ti$ nucleosynthetic yields in good agreement with gamma-ray observations of the SN1987A and Cassiopeia A supernova remnants.

We suggest, instead, that the apparent scatter in $^{44}Ca$, compared to the other Ca-isotopic ratios is again a consequence of a relatively small degree of inhomogeneous GCE. Since the synthesis of $^{44}Ca$ (as $^{44}Ti$ in an α-rich freeze-out from nuclear statistical equilibrium) is fundamentally different than that of $^{40,42,43}Ca$ (primarily O burning), one may expect that its yield may be uncorrelated with those of the other isotopes. In such a case, stochastic admixtures of supernova material into the interstellar medium might lead to more scatter in $^{44}Ca/^{40}Ca$ than the other Ca isotope ratios. As discussed in the previous section, there are substantial variations in the nucleosynthetic yields of different core-collapse supernova models, making quantitative heterogeneous modeling difficult. Nevertheless, as a first test of this idea to explain the Ca data, we considered the yields of Woosley and Heger [64]; we used these rather than the yields considered in Section 4.3 because more masses are included, potentially providing a better estimate of yield correlations. We calculated Ca δ-values for 11 bulk supernova ejecta models of mass 11-40 $M_{\odot}$. We find that the linear Pearson correlation coefficient between the $\delta^{42}Ca$ and $\delta^{43}Ca$ values for the SN ejecta is 0.37, while that for $\delta^{42}Ca$ and $\delta^{44}Ca$ is -0.18. In comparison, the correlation coefficients for these ratios for the presolar hibonite grains are 0.70 and -0.23, respectively. Therefore, one would expect that heterogeneous GCE would lead to uncorrelated (or mildly negatively correlated) $^{42}Ca/^{40}Ca$ and $^{44}Ca/^{40}Ca$ ratios, but correlated $^{42}Ca/^{40}Ca$ and $^{43}Ca/^{40}Ca$ ratios, as seen in the grain data. Again, this needs to be investigated further with modern supernova yields and GCE models, but we argue that the grain data support that a small degree of inhomogeneous GCE is needed to explain the Mg, Ti, and Ca isotopic data of M526-69.

## 5. Summary and Conclusions

We have presented Mg-Al, Si, Ca, and Ti isotopic data for an unusual presolar oxide/silicate aggregate grain, M526-69, identified in the primitive ordinary chondrite Meteorite Hills 00526. The ≈1µm aggregate consists of a Mg- and Ca-rich silicate, a smaller Al-rich oxide, and an even smaller $TiO_2$ grain. It was originally identified on the basis of large O isotopic anomalies [14], but the new multi-element data provide much more insight into the origin of the grain and constraints for the nucleosynthesis and Galactic evolution of these elements. Our main findings are:

1) M526-69 is a typical Group 2 grain in its O isotopic ratios and inferred initial $^{26}Al/^{27}Al$ ratio (Fig. 4ab). The origin of Group 2 Presolar O-rich grains is still uncertain with both low-mass and intermediate-mass AGB stars being viable, based on current stellar models and nuclear reaction rates. However, the Si isotopic ratios of the grain place it directly on the array formed by mainstream presolar SiC grains on a Si 3-isotope plot (Fig. 4d). Since IM-AGB stars are expected to exhibit substantial $^{30}Si$ excesses, this strongly indicates that the grain formed in a low-mass (<1.5 $M_{\odot}$) star. Moreover, this is most likely the case for all or most Group 2 grains, but Si data is still sparse for such grains.

2) The $^{25}Mg/^{24}Mg$, $^{29}Si/^{28}Si$, $^{30}Si/^{28}Si$, $^{42}Ca/^{40}Ca$, $^{43}Ca/^{40}Ca$, $^{47}Ti/^{48}Ti$, $^{49}Ti/^{48}Ti$, and $^{50}Ti/^{48}Ti$ ratios in M526-69 are all lower than the solar system values (Fig. 3), indicating that the parent star of the grain had lower-than-solar metallicity. Given uncertainties in GCE models, we do not provide a quantitative estimate of the star's metallicity, but note that the aggregate's Ti isotopes are similar to those of SiC Z-grains (Fig 8), thought to come from AGB stars of 1/3 to ½ solar metallicity.

3) The co-existence of Al-rich and Mg-rich phases in aggregate M526-69 allowed for an isochron-like diagram to be generated (Fig. 5) and the unambiguous determination of both the grain's initial non-radiogenic $^{26}Mg/^{24}Mg$ ratio as well as the amount of live $^{26}Al$ present in the grain at its time of formation. The initial $^{26}Mg/^{24}Mg$ ratio is slightly enriched, relative to solar, but within the range of typical, "normal" presolar silicates [32].

4) Similarly, the co-existence of Ca-rich and Ti-rich phases in the grain allow for the unambiguous determination that the small $^{44}Ca$ excess observed is not due to in situ decay of $^{44}Ti$. This proves that $^{44}Ca$ excesses in presolar grains do not necessarily point to $^{44}Ti$ and a supernova origin.

5) The Ca and Ti isotopic ratios in M526-69 reflect the initial composition of its parent star, set by GCE processes. Apart from $^{46}Ti/^{48}Ti$ and $^{44}Ca/^{40}Ca$, the rest of the Ti and Ca ratios are in good agreement with GCE trends seen in presolar SiC and/or oxide grains. Because the parent stars of SiC grains have experienced dredge-up of material that has experienced neutron capture reactions but those of oxides and silicates have not, the latter provide better constraints on the Ti-isotopic evolution of the presolar Milky Way, with much higher precision than obtained by astronomical observations.

6) The slightly enriched $^{26}Mg/^{24}Mg$, $^{46}Ti/^{48}Ti$, and $^{44}Ca/^{40}Ca$ ratios seen in M526-69, compared to the other Mg, Ca and Ti isotopic ratios, most likely reflect a small level of inhomogeneous GCE, due to stochastic variations in star formation in the Galaxy. A detailed investigation of this is beyond the scope of this work but desirable for the future, but a simple mixing model with two sets of supernova yields [58,59] supports that the ~10% shift in $^{46}Ti/^{48}Ti$ would be consistent with heterogeneous GCE. Moreover, the good correlation between $\delta^{42}Ca$ and $\delta^{43}Ca$ but slight negative correlation between $\delta^{42}Ca$ and $\delta^{44}Ca$ seen in the presolar grain data is roughly matched by correlations between predicted Ca-isotopic compositions of Type II supernovae of different masses [64]. This provides support that heterogeneous GCE is the likely explanation for the grain data and suggests that with future work the grains may provide new and unique constraints on this process in our Galaxy.

Although rare, complex aggregate multi-phase presolar grains are extremely valuable for nuclear astrophysics as they can both provide isotopic compositions for multiple elements that must be matched at a single time and place in a single star. Moreover, depending on the mineral phases that make up an aggregate, they can provide deconvolution of stable and radiogenic components for some isotopes (e.g., $^{26}Mg$ and $^{44}Ca$ studied here) not generally possible for single phases. We hope that the very interesting aggregate grain M526-69 spurs searches for more such grains as well as improved modeling of the galactic evolution of isotopic ratios with modern nucleosynthetic yield compilations.

## Acknowledgements

We dedicate this work to the late Professor Roberto Gallino, a dear friend who loved presolar grains and spent so much time trying to decode their nuclear messages. We thank Professors Adrian Brearley and Elena Dobrică for providing the MET 00526 thin section studied here, NASA for supporting this work through grant 80NSSC23K0422 to LRN, and the NSF/NASA Antarctic Search for Meteorites Program for finding this interesting meteorite.